# Planar Cu and O hole densities in high-$T_c$ cuprates determined with NMR

J. Haase[1], O. P. Sushkov[2,3], P. Horsch[3], G.V.M. Williams[4]

[1]Max-Planck-Instiute for Chemical Physics of Solids, D-01187 Dresden, Germany
[2]School of Physics, University of New South Wales, Sydney 2052, Australia
[3]Max-Planck-Institut für Festkörperforschung, D-70569 Stuttgart, Germany
[4]Industrial Research Ltd., P.O.B. 31310, Lower Hutt, New Zealand

Author to whom correspondence should be addressed:

Jürgen Haase
j.haase@physik.uni-stuttgart.de
haase@cpfs.mpg.de
Tel.: +49-351-4646-0
Fax: +49-351-4646-10

## Abstract

The electric hyperfine interaction observable in atomic spectroscopy for O and Cu ions in various configurations is used to analyze the quadrupole splitting of O and Cu nuclear magnetic resonance (NMR) in $La_{2-x}Sr_xCuO_4$ and $YBa_2Cu_3O_{6+y}$ and to determine the hole densities at both sites as a function of doping. It is found that in $La_{2-x}Sr_xCuO_4$ all doped holes (x) reside in the Cu-O plane but almost exclusively at O. For $YBa_2Cu_3O_{6+y}$ and $y<0.6$ doped holes are found at planar Cu as well as O. For $y>0.6$ further doping increases the hole content only for planar O. The phase diagram based on NMR data is presented. Further implications from the Cu A and B site in $La_{2-x}Sr_xCuO_4$ and the two planar O sites in $YBa_2Cu_3O_{6+y}$ and consequences for the phase diagram are discussed.

## Introduction

The cuprate parent compounds are Mott insulators. The Cu atoms, electronically close to a $3d^9$ configuration, form a planar square lattice with O atoms in a configuration close to $2p^6$ between them. The Cu spins order antiferromagnetically throughout the whole material. Upon doping with holes or electrons, e.g., by replacing some of the out-of-plane atoms with atoms that prefer different oxidation states, the long-range antiferromagnetism is destroyed and the materials become conductors and even superconductors at lower temperatures. For example, in $La_{2-x}Sr_xCuO_4$ $La^{3+}$ is exchanged with $Sr^{2+}$ and the common wisdom is that this increases the concentration of holes in the Cu-O plane. However, it is not quite clear whether all holes indeed reside in the plane and how they are shared between Cu and O. The situation with the in plane hole concentration and distribution in other superconducting cuprates is even less clear. Indirect chemical methods like solid solutions [1], semiempirical bond valence sums determined from structural bond lengths [2-4], or methods based on the Fermi surface topology[5] are used to determine the hole concentration. Often, the phase diagrams are just drawn by assuming a unique $T_c$ dependence on some controllable paramter that is believed to be a simple function of the hole concentration in the plane. To the best of our knowledge there are no direct physical measurements that can establish a phase diagram in terms of the hole concentration of the Cu-O plane.

A local probe like NMR that can distinguish between the various atoms in the unit cell should be able to address related questions. In fact, there have been various attempts to interpret the NMR data, the nuclear quadruole interaction for planar Cu and O, in terms of the local hole densities (see, e.g., Ref.[6] and references cited in it). This is indeed a sensible approach since the electric field gradient (EFG) at a nuclear site is very sensitive to changes in the local charge distribution. And, it is known that the planar Cu and O EFGs show a pronounced doping dependence, e.g., contrary to that of the apical O. On the experimental side these analyses used data available until 1995 and on the theoretical side they relied on atomic and cluster calculations to relate the hole concentration to the EFG tensors for Cu and O.

In the present work we perform a somewhat different analysis. First, we practically do not rely on calculations that relate the hole concentrations to the EFG components. We use data from the electric hyperfine splitting for the isolated ions. Second, we use experimental data for the parent compounds of $La_{2-x}Sr_xCuO_4$ and $YBa_2Cu_3O_{6+y}$. Altogether, this allows us to reduce the theoretical uncertainty substantially and to perform an almost model independent analysis. Due to the lack of data a complete analysis could not be performed for other materials.

The main results of the analysis, the hole densities at planar Cu and O in $La_{2-x}Sr_xCuO_4$ and $YBa_2Cu_3O_{6+y}$, are summarized in Tab. 1 under Results. We then proceed with the Data Analysis where we explain how the hole doping affects the occupation of the Cu and O orbitals. It follows a more general material specific Discussion. In Appendix 1 we calculate how the occupation of certain atomic orbitals leads to the NMR quadrupole splitting. In particular, we show how the electric hyperfine splitting observable in atomic spectroscopy can be used to remove the ambiguity in determining the radial charge distribution. In Appendix 2 we give a more detailed collection of NMR data used for the analysis.

## Results

Let us introduce in Tab. 1 the outcome of the analysis for $La_{2-x}Sr_xCuO_4$ and $YBa_2Cu_3O_{6+y}$. Next to the first column with the chemical composition we show the superconducting transition temperature $T_c$; in columns 3 to 5 three quadrupole frequencies are given. All these data are from the literature and a more complete account with references can be found in Appendix 2. Since the Cu EFG (electric field gradient at the Cu nucleus) is almost axially symmetric there is only one value to quote, ${}^{63}\nu_{Cu,c}$, the NQR frequency or the NMR splitting with the magnetic field in crystal c-direction (s. App. 1).

| Compound | $T_c$ (K) | ${}^{63}\nu_{Cu,c}$ (MHz) | ${}^{17}\nu_{Oc}$ (MHz) | ${}^{17}\nu_{Ob}$ (MHz) | $n_d$ | $n_{pc}$ | $n_{pb}$ | $\delta$ | $P_d$ | $P_p$ |
|---|---|---|---|---|---|---|---|---|---|---|
| $La_{2-x}Sr_xCuO_4$ | | | | | | | | | | |
| x = 0.0 | 0 | 33.2 | 0.147 | 0.574 | 0.780 | 0.110 | 0.110 | 0 | 0 | 0 |
| x = 0.075 | 22 | 34.2 | 0.18 | 0.6 | 0.784 | 0.137 | 0.117 | 0.058 | 0.07 | 0.93 |
| x = 0.10 | 28 | 34.6 | 0.195 | | 0.785 | 0.149 | | 0.084 | 0.06 | 0.93 |
| x = 0.15 | 37 | 35.8 | 0.215 | 0.69 | 0.794 | 0.166 | 0.153 | 0.125 | 0.11 | 0.89 |
| x = 0.20 | 30 | 36.6 | 0.245 | | 0.797 | 0.190 | | 0.177 | 0.09 | 0.90 |
| x = 0.24 | 18 | 37.4 | 0.28 | 0.81 | 0.798 | 0.219 | 0.202 | 0.236 | 0.08 | 0.92 |
| $YBa_2Cu_3O_{6+y}$ | | | | | | | | | | |
| y = 0 | 0 | 23.8 | 0.31 | 0.795 | 0.780 | 0.110 | 0.110 | 0 | 0 | 0 |
| y = 0.60 | 60 | 28.9 | 0.341 | 0.889 | 0.828 | 0.135 | 0.148 | 0.099 | 0.49 | 0.51 |
| y = 0.63 | 62 | 28.9 | 0.347 | 0.905 | 0.827 | 0.140 | 0.155 | 0.107 | 0.44 | 0.56 |
| y = 0.68 | 84 | 31.5 | 0.353 | 0.913 | 0.853 | 0.145 | 0.158 | 0.143 | 0.51 | 0.49 |
| y = 0.96 | 92 | 31.5 | 0.362 | 0.954 | 0.851 | 0.152 | 0.175 | 0.156 | 0.46 | 0.54 |
| y = 1.00 | 93 | 31.5 | 0.378 | 0.986 | 0.848 | 0.165 | 0.188 | 0.179 | 0.38 | 0.62 |
| $YBa_2Cu_4O_8$ *) | 81 | 29.72 | 0.361 | 0.926 | 0.833 | 0.152 | 0.163 | 0.136 | 0.39 | 0.61 |

Table 1: See text for explanation. *)The EFG's for $YBa_2Cu_3O_{6.0}$ were used for background subtraction.

For planar O there is a substantial asymmetry to the (traceless) EFG tensor and we show two splittings: $^{17}\nu_{Oc}$ is the value observed with the magnetic field in the direction of the crystal c-axis (this component is easily measured with aligned powder and therefore reliable and most often reported); $^{17}\nu_{Ob}$ is the value observed with the magnetic field along the Cu-O bond (it can also be determined, less reliably, from the structure of powder spectra). Since the quoted NMR studies do not measure the sign of the quadrupole interaction, we resort to positive numbers in Tab. 1.

The remaining columns 6 through 11 show data from our analysis: $n_d$ is the concentration of Cu holes with $d(x^2 - y^2)$ symmetry, $n_{p,c}, n_{p,b}$ that of oxygen holes in the $p_\sigma$ orbital as derived from $\nu_{O,c}$ and $\nu_{O,b}$, respectively. For the calculation of the remaining quantities we set $n_p \equiv n_{p,c}$. Then, $\delta = n_d + 2n_p - 1$, cf. (7), is the doping per $CuO_2$ retrieved from $n_d$ and $n_p$. $P_d$ and $P_p$ are the probabilities for the doped hole to reside on Cu and O, respectively, see (8). Our results for the total hole concentration $\delta$ are reasonably consistent with previous chemical analyes[1-4].

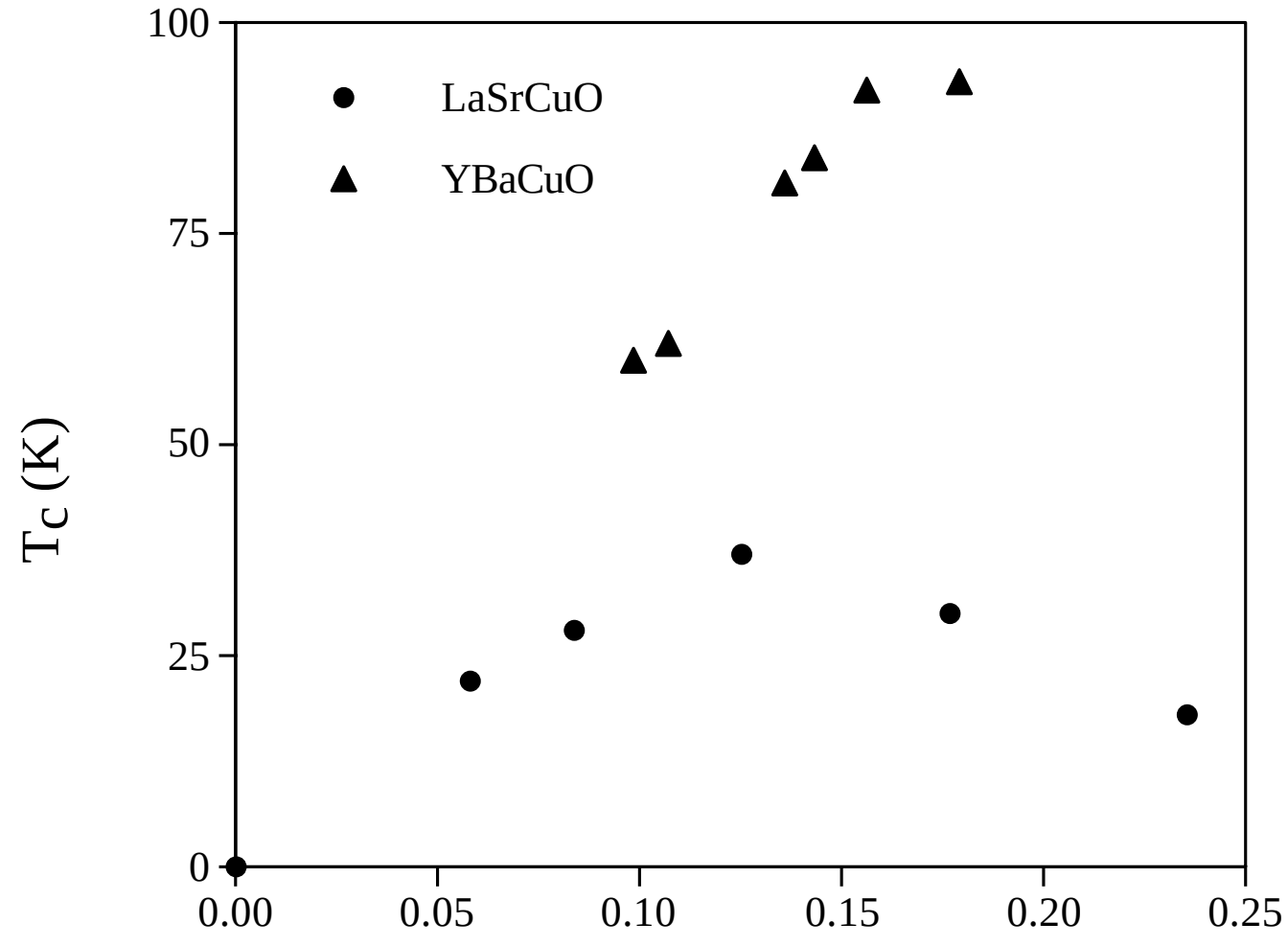

Fig. 1: Phase diagram for $La_{2-x}Sr_xCuO_4$ and $YBa_2Cu_3O_{6+y}$ as derived from NMR data shown in Tab. 1. The transition temperature $T_c$ is plotted against the hole content per $CuO_2$, given by $\delta$ as determined from NMR data.

For $La_{2-x}Sr_xCuO_4$ the Cu data refer to the so-called A-line. For $YBa_2Cu_3O_{6+y}$ there can be two oxygen lines, the mean splitting has been used. For more details see Discussion below.

## Data Analysis

We will discuss the determination of the local hole concentration for planar oxygen, first. Consider an isolated $O^{2-}$ ion. It has the electronic configuration $1s^2 2s^2 2p^6$. This is a spherically symmetric closed shell configuration and therefore the EFG at the nucleus vanishes. If one places the ion in a lattice the following effects contribute to the EFG: (a), there is a virtual or real hopping of a 2*p* oxygen electron to other ions. With some probability this creates a 2*p* hole ($1s^2 2s^2 2p^5$ configuration) that gives a substantial contribution to the EFG. (b), there is a virtual hopping of electrons from neighboring ions to unoccupied oxygen orbitals, e.g., to the 3*p* orbital. With some probability this creates the configuration $1s^2 2s^2 2p^6 3p^1$ which contributes to the EFG. (c), the electric field of distant ions has a nonzero gradient at the oxygen nucleus and hence contributes. (d), the electric field of distant ions deforms closed shells and this also creates an EFG at the oxygen nucleus.

Since a $2p$ contribution wins over that of $3p$ (due to the larger principle quantum number and a smaller population of $3p$) the above mentioned process (a) is clearly dominating. It gives the largest contribution and it is directly proportional to the hole concentration in the $2p$ orbit. All other mechanisms are expected to give smaller contributions and will practically be independent on doping. In addition, the doping dependent process should be axially symmetric. We assume that hole doping changes the occupation of $p_\sigma$ oxygen orbital only (all evidence supports this assumption, in addition, our analysis finally shows, see below, that this is indeed the case). Then, we expect that the following relation holds,

$$\nu_{O,\alpha} = q_{p\alpha} \cdot n_p + C_{O,\alpha} , \tag{1}$$

where α denotes a particular direction of the magnetic field with respect to the Cu-O plane; the concentration of holes in the $2p_\sigma$ orbital of oxygen is $n_p$; the constant $q_{p,\alpha}$ is related to the process (a) while $C_{O,\alpha}$ is due to processes (b-d). More precisely, we have to say that some contribution to $C_{O,\alpha}$ comes from the first mechanism (a), as well. For example, the oxygen $2p$ electron can virtually hop to $2p_\pi$ orbitals creating holes that contribute to the EFG. This results in some contribution that is independent of doping and therefore we absorb it into $C_{O,\alpha}$. The constants $q_{p,\alpha}$ in (1) have been calculated in App. 1 (A.31) and we can write for the oxygen quadrupole splittings with the external field in crystal c-direction, $\nu_{O,c}$, and with the field along the Cu-O bond, $\nu_{Ob}$,

$$\nu_{O,c} = q_{O,c} n_p + C_{O,c} = 1.226\, MHz \cdot n_p + C_{O,c}$$
$$\nu_{O,b} = q_{O,b} n_p + C_{O,b} = 2.453\, MHz \cdot n_p + C_{O,b} \tag{2}$$

In the parent compound the Cu hole will jump to the oxygen sites due to the strong covalency of the charge transfer insulator. The related probability amplitude $\alpha$ depends on the oxygen-copper hopping matrix element $t_{pd}$ and on the charge transfer gap $\Delta$. Values for these parameters are well established and are approximately given by $t_{pd}$ = 1.3 eV, $\Delta$ = 3.6 eV to $t_{pd}$ = 1.6 eV, $\Delta$ = 5 eV, see Ref. [7-10]. The value for the probability amplitude is[10],

$$\alpha \approx -\frac{1}{2\sqrt{2}} \sqrt{1 - \frac{1}{\sqrt{1 + 12\left(t_{pd}/\Delta\right)^2}}} \approx -0.20. \tag{3}$$

This number is closely related to the superexchange mechanism and thus to the value of the effective Heisenberg spin-spin exchange constant $J$. So a value of $\alpha \approx -0.20$ should practically be the same for all cuprates. Therefore, for the undoped compounds the hole concentration in the $2p_\sigma$ orbital of oxygen is equal to $n_p \approx 2\alpha^2 \approx 0.08$. This estimate is consistent with a more detailed numerical study[11] of the Cu-O 3-band model[12] that yields $n_p \approx 0.11$, the value we shall use in this study. Inserting this value in (A.31) gives an oxygen splitting of 0.135 MHz and 0.27 MHz for $\nu_{O,c}$ and $\nu_{O,b}$, respectively. Using these values together with (2) and the oxygen data for the parent compounds given in Tab. 1 one finds

$$\begin{array}{lll} La_2CuO_4 & C_{O,c} = 0.012\, MHz, & C_{O,b} = 0.314\, MHz \\ YBa_2Cu_3O_{6.0} & C_{O,c} = 0.175\, MHz, & C_{O,b} = 0.525\, MHz \end{array} \tag{4}$$

One could try to calculate the constants $C_{O,\alpha}$ that are anisotropic and vary between $La_{2-x}Sr_xCuO_4$ and $YBa_2Cu_3O_{6+y}$. However, such calculations require many theoretical assumptions and hence we do not want to rely on them. With the constants $C_{O,\alpha}$ determined, we can now calculate the hole content $n_p$ as a function of doping from the data in Tab. 1 using (2). We stress that the analysis based on data in c-direction is completely independent on the analysis based on data in b-direction. The values of $n_{pc}$ and $n_{pb}$ obtained from $^{17}\nu_{O,c}$ and $^{17}\nu_{O,b}$ agree very well and this confirms the present analysis. To be specific, below we will set $n_p = n_{pc}$ because data for $n_{pc}$ are more complete.

For the planar Cu the following effects contribute to the EFG at the nucleus: (a), there is a substantial concentration of $3d\left(x^2 - y^2\right)$ holes on the copper ion. This is the most important contribution to the field gradient and it is directly proportional to the hole concentration $n_d$ in this orbit. (b), there is a virtual hopping of electrons from neighboring oxygen ions to unoccupied 4p orbitals of Cu. (c), the electric field of distant ions has a nonzero gradient at the Cu nucleus and (d) electric fields of distant ions deform shells of the Cu ion. As the doping independent apical oxygen splitting in both materials shows, there is no reason to assume that the hole doping affects the Cu $3d(z^2 - r^2)$ orbital.

The processes (a) and (b) have been analyzed in previous work[10], here we follow this analysis. Similar to oxygen, the contribution of process (a) depends on doping. The higher orbital momentum of $d$ as compared to $p$ electrons makes contribution (b) important despite the larger principal quantum number of the 4p state.

With the magnetic field perpendicular to the plane the quadrupole splitting due to the $3d(x^2 - y^2)$ hole is given by (A.32). The oxygen 2p electrons also contribute to the Cu EFG. In the vicinity of Cu, the oxygen $2p_\sigma$ wave function can be expanded in the basis of the Cu orbitals: $|2p_\sigma\rangle \rightarrow \beta|4p\rangle$. The expansion coefficient β has been calculated previously[10], $\beta^2 \approx 0.40$. Every electron on the Cu 4p orbital gives a contribution to the EFG according to (A.33). There are $\left(8 - 4n_p\right)$ electrons on the four nearest oxygen atoms and each of them penetrates to the $4p$ orbital with probability $\beta^2$. This completes the expression for the quadruole frequency for Cu,

$$\nu_{Cu,c} = 94.3MHz \cdot n_d - 14.2MHz \cdot \beta^2\left(8 - 4n_p\right) + C_{Cu,c} \quad (5)$$

where the last term $C_{Cu,c}$ is due to the mechanisms (c) and (d) and can be different for $La_{2-x}Sr_xCuO_4$ and $YBa_2Cu_3O_{6+y}$. Thus far, we have determined the first two terms of (5). The first term in this equation is very reliable. It is based completely on experimental data. The second term depends on $\beta^2 \approx 0.40$ which is a theoretical quantity. However, the calculation is simple and we believe that the second term is also reliable. Moreover, in principle one does not need to rely on a calculation of $\beta^2$. It can be determined from NMR data of the parent material: since the coefficients $C_{Cu,c}$ in (5) turn out to be much smaller than the other terms, cf. (6), they can be neglected in first approximation. Then by applying (5) to the parent material one immediately finds that $\beta^2 \approx 0.4$. The first two terms in (5) should be the same for all cuprates. The last term $C_{Cu,c}$, as mentioned earlier, depends on the details of the lattice structure. Similar to the oxygen EFG we do not see a way to obtain it reliably from theory. Fortunately, we can find values for $C_{Cu,c}$ from NMR data for the parent compounds. For the undoped materials $n_p = 2\alpha^2 = 0.11$ and thus $n_d = 1 - 4\alpha^2 = 0.78$. Using this value together with (5) and the Cu frequencies for the parent materials from Tab. 1 we obtain

$$La_2CuO_4 \quad C_{Cu,c} = +2.59 MHz$$
$$YBa_2Cu_3O_{6.0} \quad C_{Cu,c} = -6.81 MHz \quad (6)$$

Various experimental data points[13] for planar Cu in $YBa_2Cu_3O_{6+y}$ show that these constants are quite reliable.

Now, using $C_{Cu,c}$ and (5) we have calculated $n_d$ for all materials in Tab. 1. Since we determined $n_d$ and $n_p$ we can calculate the doping in the plane based on NMR data,

$$\delta = n_d + 2n_p - 1 \quad (7)$$

(to be precise, we used $n_p = n_{pc}$). The values for $\delta$ are listed in Tab. 1, as well. In order to view the distribution of holes in the plane more easily let us represent the densities in the following form,

$$n_d = n_{d0} + P_d\delta$$
$$n_p = n_{p0} + \frac{1}{2}P_p\delta \quad (8)$$

where $n_{d0}$ and $n_{p0}$ are densities without doping and $P_d$ and $P_p$ are probabilities for the doped hole to go to the copper or the oxygen ion, respectively. There is a coefficient 1/2 in the second equation because there are two oxygen atoms in the unit cell. The values for $P_d$ and $P_p$ obtained from (8) are also listed in Tab. 1.

## Discussion

*$La_{2-x}Sr_xCuO_4$*

In the parent compound, NMR finds one planar O site and one Cu site [14]. Upon doping, however, a second Cu site (B) appears [15, 16]. The NMR intensity of site B grows linearly with x and corresponds to about 13% of all Cu sites for x = 0.15. It has been suggested in the early work[15, 16] that B represents Cu bonded via apical O to Sr as opposed to La. More recently, another explanation has been brought forward[17] that sees the Cu B site caused by the 4 Cu next to a localized Cu hole. However, the NMR signals of the hole site and the nearest O have not been observed. First principle calculations[18] do not agree with this view, and, since both Cu sites A and B show very similar NMR linewidths [19] the explanation that site B is close to a localized hole can be dismissed. Therefore, the main site A quadrupole splitting was used for our analysis in Tab. 1. Nevertheless, the Cu B site shows a slower nuclear relaxation as compared with the A site[20] which must be due to differences in the magnetic hyperfine scenario. In addition, the experimentally observed dependence of the quadrupole splitting on x, which is linear for both Cu sites, has a much smaller slope for the B site where it is only about 1/6 of that of the A site (~19.3 MHz/x) [21]. This shows that the B site Cu does not see the same average hole density increase upon doping that determines the A site splitting. All these experimental data are consistent with the Cu B site being near a higher hole density. While a slight increase in the hole concentration on Cu itself does not affect the overall doping level, a scenario where the holes reside on O has consequences for the phase diagram. Consider x = 0.05. Here the B site has a splitting of 38.5 MHz[21]. This corresponds with (5) and (6) to a local hole concentration per unit cell of about 0.35, assuming the same constant $C_{Cu,c}$ for the B site. Since 5% of the Cu atoms experience this doping level we have 95% of the Cu atoms experiencing a local doping level of only $\delta = 0.034$, a value close to where $T_N$ vanishes. Since additional structural changes might influence the numbers, the so-called spin-glass region in the $La_{2-x}Sr_xCuO_4$ phase diagram could be a consequence of the fact that the Cu B site holds most of the doped carriers and prevents superconductivity in this region, cf. also Fig. 1. However, in such a scenario one would expect that the planar O surrounding the B site should have a splitting of 0.215 MHz for x = 0.05, a resonance that might be detectable despite the large NMR linewidths.

*$YBa_2Cu_3O_{6+y}$*

In this material NMR finds one planar Cu line but there can be two planar O signals of similar intensity[22, 23] whose appearance depends on doping. Originally, these two signals were explained with the orthorhombic structure, however, this interpretation was challenged[24] recently since the extent of the splitting and its doping dependence does not agree with what one expects from the doping dependence of the orthorhombic distortion. Alternatively, the splitting might indicate the presence of a commensurate charge density wave. Unfortunately, there are not enough data available for a firm conclusion. For $YBa_2Cu_3O_{6+y}$, y = 1.0 the difference in the splitting is $\Delta\nu_{O,c} = 0.018 MHz$, cf. Appendix 2. Thus, the hole variation between the two oxygen sites could easily amount to 0.015 which is 9% of 0.165, the average hole density on O found in Tab. 1. Since a single, narrow (~ 98 kHz linewidth[25]) Cu line is observed, the charge variation for Cu has to average. As a consequence a commensurate one-dimensional charge density variation that runs parallel or at a 45 degree angle with respect to the Cu-O bond explains the data if the charge density alternates for subsequent lines of oxygen. The relation of such a modulation to the direction of the chains above the Cu-O plane is not known. Further experiments are necessary.

Our analysis of the hole concentrations for both materials was based on the average NMR splittings. While some details, the two Cu sites in $La_{2-x}Sr_xCuO_4$ and the two O sites in $YBa_2Cu_3O_{6+y}$, have been discussed, we have to mention that basically all cuprates show excessive quadrupolar as well as magnetic linewidths (distributions of splittings and shifts). In fact, it has been shown[26] that both distributions are intimately linked, a fact that is not well understood. A more detailed account of linewidths and consequences for the hole distribution will be addressed in a separate report.

If we confer with Tab. 1 we see that for $La_{2-x}Sr_xCuO_4$ the concentration of holes in Cu $3d\left(x^2-y^2\right)$ is almost independent on doping, i.e., it remains close to the value given by the virtual hopping in the parent material. Almost all doped holes as inferred from the value of $x$ go to oxygen. The situation is somewhat different for $YBa_2Cu_3O_{6+y}$. Here we see that with a probability of about 30-40% the doped holes are located on Cu, with a probability of 60-70% they are found on O. This agrees reasonably with the predictions of the three band model with standard parameters[7-9]. The data in Tab. 1 also show that the total hole concentration in the Cu-O plane $\delta$ for $La_{2-x}Sr_xCuO_4$ and $YBa_2Cu_3O_{6+y}$ is proportional to x and y, respectively, with slopes of 0.97 and 0.18. This shows that per Sr almost a full hole is introduced into the plane, whereas, in $YBa_2Cu_3O_{6+y}$ only about 1/5 of a hole goes to the plane per added O. From Fig. 1 it appears that at optimal doping the hole concentration in $YBa_2Cu_3O_{6+y}$ is substantially larger. A similar trend seems likely from the limited data available for other cuprates: at optimal doping $T_c$ increases with the planar oxygen splitting indicating a larger hole content. In addition to the data presented is seems worth noting that the quadrupole splittings for the apical oxygens, that are independent on doping, are very different. For the parent materials one finds that the axially symmetric EFG corresponds to oxygen hole concentrations of 0.08 and 0.48 for $La_{2-x}Sr_xCuO_4$ and $YBa_2Cu_3O_{6+y}$, respectively.

To conclude, we have introduced a new approach for the determination of the local hole densities at planar Cu and O directly from the NMR. The electric hyperfine interactions for the ions in the various oxydation states from atomic spectroscopy were analyzed and used to determine the NMR splittings for such states. In addition, the NMR data for the parent compounds were used subtract doping independent contributions to the electric field gradient. We were thus able to give an unbiased account of the actual hole distribution upon doping the Cu-O plane for two materials where sufficient NMR data are available. From the results it was possible to independently construct a phase diagram based on the literature data on $T_c$ and the NMR splittings alone. We found that in $La_{2-x}Sr_xCuO_4$ almost all doped holes go into the Cu-O plane but reside almost exclusively on O. The hole content for $YBa_2Cu_3O_{6+y}$ could be determined as well, removing the ambiguity that stems from the presence of the chains. Here, initially some holes go to Cu whereas for y > 0.6 additional holes only go to the O. As further NMR data become available for other materials the same approach can be used to explore the hole doping that contains significant clues for the understanding of the materials.

## Acknowledgement

We would like to thank B. Keimer, G. Khaliullin, C. Bernhard, and A. Bussmann-Holder for valuable discussions, and J. Roos, M. Mali for communicating the $^{17}O$ NMR results for $YBa_2Cu_3O_6$.

## Appendix 1: Relations between nuclear quadrupole effects in atomic spectroscopy and NMR

The electric quadrupole interaction of a nuclear charge density with that of its surrounding electrons can be written[27] in terms of spherical harmonics $Y_{\mathrm{lm}}$ as

$$H_Q = \frac{4\pi}{5}\int_{\tau_e}\int_{\tau_n} \rho_e(\mathbf{r}_e)\rho_n(\mathbf{r}_n)\frac{r_n^2}{r_e^3}\sum_{q=-2}^{2}(-1)^q Y_{2-q}(\theta_n,\phi_n)Y_{2q}(\theta_e,\phi_e)d\tau_n d\tau_e, \tag{A.1}$$

where the integrals are taken with respect to the nuclear (n) and electronic (e) coordinates. It is useful to resort to irreducible tensors by writing

$$\begin{aligned} H_Q &= \sum_{q=-2}^{2}(-1)^q\sqrt{\frac{4\pi}{5}}\int \rho_n r_n^2 Y_{2-q}(\theta_n,\phi_n)d\tau_n \cdot \sqrt{\frac{4\pi}{5}}\int\frac{\rho_e}{r_e^3}Y_{2q}(\theta_e,\phi_e)d\tau_e \\ H_Q &= \sum_{q=-2}^{+2}(-1)^q Q_{2-q}V_{2q},\ \text{ e.g.,}\quad V_{20} = \sqrt{\frac{4\pi}{5}}\int\frac{\rho_e}{r_e^3}Y_{20}(\theta_e,\phi_e)d\tau_e \end{aligned} \tag{A.2}$$

One typically introduces the nuclear electric quadrupole moment in terms of the nuclear spin $I$ and the magnetic quantum number $m_I$,

$$eQ = 2\langle I\, m_I = I|Q_{20}|I\, m_I = I\rangle. \tag{A.3}$$

With the Wigner-Eckart theorem one writes in standard notation,

$$\langle I\, m_I|Q_{20}|I\, m_I\rangle = (-1)^{I-m}\begin{pmatrix} I & 2 & I \\ -m & 0 & m\end{pmatrix}\langle I\|\mathbf{Q}_2\|I\rangle. \tag{A.4}$$

For a free ion the total angular momentum $\mathbf{F} = \mathbf{J}+\mathbf{I} = \mathbf{L}+\mathbf{S}+\mathbf{I}$ is a good quantum number and by using the Wigner-Eckart theorem for irreducible tensor products one calculates the energy as follows

$$\langle Fm_F JI|H_Q|Fm_F JI\rangle = (-1)^{F+J+I}\begin{Bmatrix} J & I & F \\ I & J & 2\end{Bmatrix}\langle I\|\mathbf{Q}_2\|I\rangle\langle J\|\mathbf{V}_2\|J\rangle. \tag{A.5}$$

In analogy to (A.3) one defines the electric field gradient, $eq \equiv \langle V_{zz}\rangle$, in terms of the total electron angular momentum,

$$eq = 2\langle J\, m_J = J|V_{20}|J\, m_J = J\rangle. \tag{A.6}$$

Then, one finds[27] for the reduced matrix elements in (A.5)

$$\langle J\|\mathbf{V}_2\|J\rangle = \frac{\sqrt{(2J+3)(J+1)(2J+1)}}{2\sqrt{J(2J-1)}}eq,\ \ \langle I\|\mathbf{Q}_2\|I\rangle = \frac{\sqrt{(2I+3)(I+1)(2I+1)}}{2\sqrt{I(2I-1)}}eQ \tag{A.7}$$

With $X = I(I+1)+J(J+1)-F(F+1)$ one obtains for the quadrupole energy

$$\langle Fm_F JI|H_Q|Fm_F JI\rangle = \frac{e^2qQ\left[\frac{3}{4}X(X+1)-J(J+1)I(I+1)\right]}{2I(2I-1)J(2J-1)}. \tag{A.8}$$

In atomic spectroscopy one defines the quadrupole coupling constant $B_J$ (in frequency units, $h$ is Planck's constant) as

$$B_J \cdot h = e^2 qQ \equiv 2eQ\left\langle J\,J\left|V_{20}\right|J\,J\right\rangle = \frac{2eQ}{h}\sqrt{\frac{4\pi}{5}}\int \frac{\rho_e}{r_e^3}\left\langle JJ\left|Y_{20}\right|JJ\right\rangle dr_e\,. \qquad \text{(A.9)}$$

For a particular ionic state, ${}^{2S+1}L_J$, one can expand the total angular momentum eigenstate in terms of those of the orbital moment $L$ and spin $S$ using the Clebsch-Gordan coupling coefficients. For example, by setting $\left\langle r^{-3}\right\rangle = \frac{1}{-e}\int \frac{\rho_e}{r_e} dr_e \equiv \int \frac{\psi_e^2}{r_e} dr_e$ one calculates for the atomic states ${}^{2S+1}L_J$ the following constants $B_J$,

$$B_{3/2}\left({}^2P_{3/2}\right) = +\frac{2}{5}e^2 Q\left\langle r^{-3}\right\rangle, \qquad \text{(A.10)}$$

$$B_{3/2}\left({}^2D_{3/2}\right) = +\frac{2}{5}e^2 Q\left\langle r^{-3}\right\rangle,\ B_{5/2}\left({}^2D_{5/2}\right) = +\frac{4}{7}e^2 Q\left\langle r^{-3}\right\rangle. \qquad \text{(A.11)}$$

In solids it is convenient to write (A.5) as

$$\left\langle I,m_I,\psi_e\left|H_Q\right|I,m_I,\psi_e\right\rangle = \sum_{q=-2}^{+2}(-1)^q\left\langle I,m_I\left|Q_{2-q}\right|I,m_I\right\rangle\left\langle \psi_e\left|V_{2q}\right|\psi_e\right\rangle \qquad \text{(A.12)}$$

where $\psi_e$ is the electronic wave function. With the Wigner-Eckart theorem one sees that only the term with $q$ = 0 contributes,

$$\left\langle I,m_I,\psi_e\left|H_Q\right|I,m_I,\psi_e\right\rangle = \frac{3m^2 - I(I+1)}{2I(2I-1)}eQ\cdot\left\langle \psi_e\left|V_{20}\right|\psi_e\right\rangle. \qquad \text{(A.13)}$$

By calling the nuclear quantization axis the z-axis we may introduce the effective Hamiltonian,

$$H_{Qeff} = \frac{3I_z^2 - I(I+1)}{2I(2I-1)}eQ\cdot\left\langle \psi_e\left|V_{20}\right|\psi_e\right\rangle. \qquad \text{(A.14)}$$

For a wavefunction that is a product of an angular part and a radial part we can write

$$\left\langle \psi_e\left|V_{20}\right|\psi_e\right\rangle = -e\sqrt{\frac{4\pi}{5}}\left\langle \psi_e(\theta,\phi)\left|Y_{20}\right|\psi_e(\theta,\phi)\right\rangle\cdot\left\langle r^{-3}\right\rangle \qquad \text{(A.15)}$$

In particular, we can easily calculate the field gradient due to an atomic orbital. Writing,

$$H_{Qeff} = \frac{3I_z^2 - I(I+1)}{2I(2I-1)}e^2 Q\cdot f_{z,lm}\left\langle r^{-3}\right\rangle_{lm} \qquad \text{(A.16)}$$

we find the well known numbers (that add to zero for closed shells),

$$\begin{aligned} &\text{p-orbitals: } f_{z,p_z} = -\frac{2}{5},\ f_{z,p_x} = +\frac{1}{5},\ f_{z,p_y} = +\frac{1}{5} \\ &\text{d-orbitals: } f_{z,z^2-r^2} = -\frac{2}{7},\ f_{z,x^2-y^2} = +\frac{2}{7},\ f_{z,zx} = -\frac{1}{7},\ f_{z,yz} = -\frac{1}{7},\ f_{z,xy} = +\frac{2}{7} \end{aligned} \qquad \text{(A.17)}$$

In general in a solid the electronic wavefunction can not be written as a product of an angular and radial part due to crystal field effects. In such a case the electric field gradient may be asymmetric in the principle axis system.

A strong magnetic field $B_z$ in z-direction causes nuclear Zeeman interaction, expressed by the Hamiltonian

$$H_L = \hbar\omega_L I_z \tag{A.18}$$

where $\omega_L = -\gamma B_z$ is the Larmor frequency. To leading order[28], the Hamiltonian is now

$$H = \hbar\omega_L I_z + \frac{3I_z^2 - I(I+1)}{2I(2I-1)} eQ \cdot \frac{1}{2}\langle V_{zz}\rangle . \tag{A.19}$$

The component $\langle V_{zz}\rangle$ depends on the relative orientation of the nuclear quantization axis and the corresponding symmetric tensor is usually expressed in terms of the principle axis components $V_{XX}, V_{YY}, V_{ZZ}$, or in terms of the largest component and its asymmetry parameter,

$$\eta = \frac{V_{XX} - V_{YY}}{V_{ZZ}},\ |V_{ZZ}| \geq |V_{YY}| \geq |V_{XX}| . \tag{A.20}$$

In the NMR literature[28] one therefore writes

$$H = \hbar\omega_L I_z + \frac{3I_z^2 - I(I+1)}{2I(2I-1)} \cdot \frac{1}{2} eQV_{ZZ} \left\{ \frac{3\cos^2\theta - 1}{2} + \frac{\eta}{2}\sin^2\theta\cos 2\phi \right\}, \tag{A.21}$$

and defines the quadrupole and angular dependent quadrupole frequency by

$$\omega_Q = \frac{3eQV_{ZZ}}{2I(2I-1)} \equiv \frac{3e^2qQ}{2I(2I-1)} \tag{A.22}$$

$$\omega_Q^{'}(\theta,\phi) = \omega_Q \left\{ \frac{3\cos^2\theta - 1}{2} + \frac{\eta}{2}\sin^2\theta\cos 2\phi \right\}. \tag{A.23}$$

This leads to a simple first-order Hamiltonian

$$H = \hbar\omega_L I_z + \frac{\hbar\omega_Q^{'}}{6}\left[3I_z^2 - I(I+1)\right]. \tag{A.24}$$

For the energy of the (dipole allowed) transitions it follows,

$$\frac{E_{m+1} - E_m}{\hbar} = \omega_L + \left( m + \frac{1}{2} \right)\omega_Q^{'} . \tag{A.25}$$

With the field axis along the principal Z-axis we see that the quadrupole splitting between the central NMR line and the first satellite is just the quadrupole frequency. The principle components can also be written as $\left(V_{ZZ}, V_{YY}, V_{XX}\right) = \left(1, -\frac{1+\eta}{2}, -\frac{1-\eta}{2}\right)$. Without going into more details, since Cu can also be measured in zero field, the measured resonance frequency in NQR $\omega_{NQR} = \omega_Q\sqrt{1+\eta^2/3}$ is very close to $\omega_Q$, as well, since $\eta \approx 0$ for Cu in the Cu-O plane.

We now would like to point out that it is possible to relate the data from atomic spectroscopy with the quadrupole splitting observed in NMR experiments. We define the latter as the frequency difference between the central line and that of the first satellite,

$$\omega = \omega_Q^{'} \qquad \text{(A.26)}$$

If the magnetic field is perpendicular to the Cu-O plane we have for the splitting

$$\omega = \frac{3e^2Q}{2I(2I-1)} \cdot 2 f_{z,lm} \langle r^{-3} \rangle_l \qquad \text{(A.27)}$$

*Planar oxygen-17* has I = 5/2 and in $2p^5$ configuration we have with (A.10) in frequency units

$$\nu_O = \frac{3}{4} B_{3/2} f_{z,lm}. \qquad \text{(A.28)}$$

With the hole in the Cu-O $p_\sigma$ orbital we have for the splitting with the field perpendicular and in the bond direction,

$$\nu_{O,c} = \frac{3}{20} B_{3/2}, \quad \nu_{O,b} = \frac{3}{10} B_{3/2} \qquad \text{(A.29)}$$

respectively. Although, we are not aware of experimental data for $O^-$, there is a recent detailed atomic many-body calculation[29] of $B_{3/2}$. According to this study the constant $B_{3/2} = 8.174\,MHz$ for the ${}^2P_{3/2}$ state. It follows

$$\nu_{O,c} = 1.226\,MHz, \quad \nu_{O,b} = 2.452\,MHz. \qquad \text{(A.30)}$$

Since there is no splitting for a $2p^6$ configuration we expect the splitting to be proportional to the hole density $n_p$ in the oxygen p orbital,

$$\nu_{O,\alpha} = q_{p_\alpha} n_p, \quad q_{p_c} = 1.226\,MHz, \; q_{p_b} = 2.452\,MHz \qquad \text{(A.31)}$$

*Planar $^{63}$Cu* (I=3/2) in the $3d^9$ configuration has two ionic states, ${}^2D_{3/2}, {}^2D_{5/2}$ that can be used for determining the coupling constant [30-32]. However, the literature shows that $B_{3/2} = 138 \pm 8\,MHz,\; B_{5/2} = 188 \pm 3\,MHz$. With the help of (A.11) we find $e^2Q\langle r^{-3}\rangle = -\frac{5}{2} B_{3/2} = 345 MHz, \;\; e^2Q\langle r^{-3}\rangle = \frac{7}{4} B_{5/2} = 329 MHz$, i.e., slightly different values. We average the two values based on the error bars and for the magnetic field in c-direction ( $f_{z,x^2-y^2} = 2/7$) and obtain

$$\nu_{Cu,c}\left(3d^9\right) = 94.30 \pm 1.45 MHz. \qquad \text{(A.32)}$$

For planar $^{63}$Cu in the $3d^{10}4p$ configuration there are also spectroscopic data available[30-32], $B_{3/2} = -28.4 \pm 0.4\,MHz$. With (A.10) and (A.17),

$$\nu_{Cu,c}\left(4p\right) = -14.2 \pm 0.2\,MHz. \qquad \text{(A.33)}$$

Since one often finds values for $\langle r^{-3}\rangle$ in the literature we give the values that follow from the spectroscopic data reported above. With nuclear quadrupole moments of $^{17}Q = -0.026\; 10^{-28}\; m^2$ and $^{63}Q = -0.211\; 10^{-28}\; m^2$ [33] one has from (A.10), (A.11) in SI units

$$\left\langle a_B^3 r^{-3} \right\rangle_{3/2} = \frac{4\pi\varepsilon_0 h a_B^3}{e^2 \cdot Q} \frac{2}{5} B_{3/2}, \ \left\langle a_B^3 r^{-3} \right\rangle_{5/2} = \frac{4\pi\varepsilon_0 h a_B^3}{e^2 \cdot Q} \frac{7}{4} B_{5/2}. \tag{A.34}$$

This gives with the spectroscopic constants from above the following numbers,

$$\begin{aligned} &^{17}O: \quad \left\langle r^{-3} \right\rangle_{2p} = +3.345\, a.u. \\ &^{63}Cu: \ \left\langle r^{-3} \right\rangle_{3d} = +6.959 \ldots + 6.636\, a.u., \ \left\langle r^{-3} \right\rangle_{4p} = -1.432\, a.u. \end{aligned} \tag{A.35}$$

## Appendix 2: List of experimental data with references

| Compound | $T_c$ (K) | $^{63}\nu_{Cu,c}$ (MHz) | $^{17}\nu_{O,c}$ (MHz) | $^{17}\nu_{O,b}$ (MHz) | $^{17}\nu_{O,a}$ (MHz) |
|---|---|---|---|---|---|
| $La_2CuO_4$ | 0 | 33.2 [34] | -0.147 [14] | 0.574 | -0.427 |
| $La_{1.925}Sr_{0.075}CuO_4$ | 22 | 34.2 [35] | -0.18 [35] | 0.60 | -0.42 |
| $La_{1.90}Sr_{0.10}CuO_4$ | 35 | 34.6 [25] | -0.195 [25] | | |
| $La_{1.85}Sr_{0.15}CuO_4$ | 38 | 35.8 [35] | -0.215 [25] | 0.69 | |
| $La_{1.80}Sr_{0.20}CuO_4$ | 36 | 36.6 [25] | -0.245 [25] | | |
| $La_{1.76}Sr_{0.24}CuO_4$ | 18 | 37.4 [35] | -0.28 [35] | 0.81 | -0.53 |
| | | | | | |
| $YBa_2Cu_3O_6$ | 0 | 23.8 [13, 36, 37] | -0.31 [38] | 0.795 | -0.485 |
| $YBa_2Cu_3O_{6.60}$ | 60 | 28.9 [39] | -0.341 [39] | 0.889 | -0.544 |
| $YBa_2Cu_3O_{6.63}$ | 62 | 28.9 [40] | -0.347 [40] | 0.905 | -0.557 |
| $YBa_2Cu_3O_{6.8}$ | 84 | 31.5 [39] | -0.353 [39] | 0.913 | -0.559 |
| $YBa_2Cu_3O_{6.96}$ | 92 | 31.5 [39, 41] | -0.362 [39] | 0.954 | -0.584 |
| $YBa_2Cu_3O_7$ | 93 | 31.5 [25] | -0.387, -0.369 [22, 42] | 0.986, 0.966 | -0.598 |
| $YBa_2Cu_4O_8$ | 81 | 29.72 [43] | -0.365, -0.357 [23] | 0.927, 0.925 | -0.562, -0.568 |